\documentclass[aps,prc,amsmath,amssymb]{revtex4}

\usepackage{graphicx}% Include figure files
\usepackage{dcolumn}% Align table columns on decimal point
\usepackage{bm}% bold math

\begin{document}

\preprint{XVIII Symposium on Nucl. Phys., Cocoyoc, Jan. 4-7 2005 / Rev. Mex. F\'{\i}s.}

\title{IBM-2 configuration mixing and its geometric interpretation 
for germanium isotopes}

\author{E. Padilla-Rodal$^{1,2}$}%
\author{O. Casta\~nos$^{1}$}%
\author{R. Bijker$^{1}$}%
\author{A. Galindo-Uribarri$^{2}$}%

\affiliation{$^{1}$Instituto de Ciencias Nucleares, UNAM, 04510 M\'exico, D.F., M\'exico}%
\affiliation{$^{2}$Physics Division, Oak Ridge National Laboratory, Oak Ridge,Tennessee 37831}%

\begin{abstract}

The low energy spectra, electric quadrupole transitions, and quadrupole moments for the germanium isotopes are determined in the formalism of the IBM-2 with configuration mixing. These calculated observables reproduce well the available experimental information including the newly obtained data for radioactive neutron-rich $^{78,80,82}$Ge isotopes. Using a matrix formulation, a geometric interpretation of the model was established. The two energy surfaces determined after mixing, carry information about the deformation parameters of the nucleus. For the even-even Ge isotopes the obtained results are consistent with the shape transition that takes place around the neutron number $N=\nobreak40$. 


Los niveles de baja energ\'{\i}a, las transiciones cuadrupolares el\'ectricas y los momentos cuadrupolares de los is\'otopos de germanio son determinados en el formalismo del IBM-2 con mezcla de configuraciones. Las observables calculadas reproducen bien la informaci\'on experimental disponible incluyendo datos obtenidos recientemente para los is\'otopos radiactivos con exceso de neutrones $^{78,80,82}$Ge. Utilizando una formulaci\'on matricial, se estableci\'o una interpretaci\'on geom\'etrica del modelo. Las dos superficies de energ\'{\i}a determinadas despu\'es de la mezcla, contienen informaci\'on acerca de los par\'ametros de deformaci\'on del n\'ucleo. Los resultados obtenidos para los is\'otopos par-par de Ge son consistentes con la transici\'on de fase que ocurre alrededor del n\'umero de neutrones $N=40$.
\end{abstract}


\keywords{Modelo de bosones interactuantes; mezcla de monfiguraciones; transici\'on de forma; is\'otopos de germanio}

\maketitle

\section{\label{sec:1}Introduction}

Recent results on Coulomb excitation experiments of radioactive neutron-rich Ge isotopes at the Holifield Radioactive Ion Beam Facility allowed the study of the systematic trend of $B(E2; 0_1^+ \rightarrow 2_1^+)$ between the sub-shell closure at $N=40$ and the major-shell $N=50$~\cite{us}. The new information on the $E2$ transition strengths constitutes a stringent test for the nuclear models \cite{us,lisetskiy} and has motivated us to revisit the use of the Interacting Boson Model (IBM) for these isotopes. Previous work \cite{duval-goutte}, using a version of the IBM-2 with configuration mixing, has shown that a good description of the stable germanium nuclei can be obtained. In the present work we apply the standard, two-particle two-hole, IBM-2 with configuration mixing~\cite{duval} to the stable nuclei and extrapolate the model predictions to the recently explored radioactive neutron-rich isotopes $^{78, 80}$Ge and the single-closed shell nucleus $^{82}$Ge.

The irregular neutron-dependence of important observables such as the excitation energy of the $0_2^+$ states, the relative values of the $B(E2)'$s and the population cross sections in two-neutron-transfer reactions \cite{vergnes1978} have suggested that a structural change takes place around $N=40$ for Ge isotopes. In combination with the measurement of the electric quadrupole moments associated with the $2_1^+$ and $2_2^+$ states \cite{lecomte, toh}, this experimental data has been taken as evidence of a shape transition and the coexistence of two different kinds of deformations for this isotopic chain \cite{lecomte2}.

For many years several theoretical mechanisms have been proposed to explain these phenomena simultaneously in a consistent way. For example, in the early seventies the variation of the $0_2^+$ excitation energies was explained under the assumption of a second minima in the potential energy surface~\cite{gneuss1971}. However the success of this description was limited as the excited states were not well reproduced. Investigations of the nuclear structure with the dynamic deformation theory~\cite{kumar1} were also performed leading to the determination of potential energy surfaces and energy levels of the Ge isotopes. Although these calculations were not able to predict correctly the $2_2^+$ state for the $^{72}$Ge, the results implied that the Ge nuclei were very soft and present an oblate-prolate shape phase transition~\cite{kumar2}. Another relevant work that uses a boson Hamiltonian to describe the quadrupole degrees of freedom for the Ge isotopes, is the study based on the coupling of pairing and collective quadrupole vibrational modes~\cite{weeks} through a boson expansion procedure~\cite{tamura}. This formalism describe successfully many features of the Ge isotopes, although it had some difficulties in fitting some of the two-nucleon transfer cross sections.     

\section{\label{sec:2}IBM-2 with configuration mixing for Ge isotopes}

Under the assumption that the $0_2^+$ states in the germanium isotopes arise from an intruder configuration, in this contribution we reconsider the formalism of the IBM-2 with configuration mixing to describe the nuclear structure of these nuclei. In the IBM-2 the nucleus is modeled as a system of two types of interacting bosons, proton- and neutron-bosons, that can have angular momentum and parity $J^P = 0^+, 2^+$ and are denoted by the creation(annihilation) operators $s^\dagger_\rho$($s_\rho$), and $d^\dagger_\rho$($d_\rho$), respectively, where $\rho = \pi$ indicates protons and $\rho = \nu$ is used for neutrons.

The mixing calculation consists of first describing the general features of the two configurations in terms of two different IBM-2 calculations and then combining these two results using a mixing operator. Each configuration is described using a Hamiltonian of the form
	\begin{equation}
	H = \varepsilon n_d + \kappa Q_\pi \cdot Q_\nu + M_{\pi \nu} \, , 
	\end{equation}
where $n_d =\sum_{\mu,\, \rho}( d^\dagger_{\mu \rho} d_{\mu \rho})$ denotes the number operator of $d$-bosons, $Q_\rho$ represents the quadrupole operator for protons and neutrons 
	\begin{equation}
	Q_\rho = (s^\dagger_\rho \widetilde d_\rho + d^\dagger_\rho s_\rho)^{(2)} + 
	\chi_\rho (d^\dagger_\rho \widetilde d_\rho)^{(2)} \, ,
	\label{quadrupole}
	\end{equation}
and $M_{\pi \nu}$ is the Majorana interaction 
	\begin{equation}
	M_{\pi \nu} = \xi_2 \
	( s^\dagger_\pi \, d^\dagger_\nu - d^\dagger_\pi \, s^\dagger_\nu )^{(2)} \ 
	( s_\pi \, \widetilde d_\nu - \widetilde d_\pi \, s_\nu )^{(2)} 
	 +  \sum_{K = 1,3} \ \xi_K \ (d^\dagger_\pi \, d^\dagger_\nu)^{(K)}
	(\widetilde d_\pi \, \widetilde d_\nu)^{(K)} \, .
	\end{equation}
The two Hamiltonians are diagonalized independently in its appropriate space. The active model space for protons in the normal configuration consists of two proton-bosons, whereas the intruder space is conformed of four proton-bosons, one boson-hole in the $20$-$28$ shell and three boson-particles in the $28$-$50$ shell. The mixing Hamiltonian that connects this two configurations does not conserve the number of bosons and is given by
	\begin{equation}
	H_{\rm mix} = \alpha_0 (s^\dagger_\pi  s^\dagger_\pi + s_\pi  s_\pi)
	+ \alpha_2 (d^\dagger_\pi \times d^\dagger_\pi +
	\widetilde d_\pi \times \widetilde d_\pi)^{(0)} \, . \label{hmix}
	\end{equation}
A third parameter, $\Delta$, is needed in order to specify the unperturbed energy required to excite two protons across the closed shell~\cite{pittel}. Using the eigenfunctions of the two separate configurations one forms the matrix elements of $H_{\rm mix}$. The final wave functions are obtained from the diagonalization of the resulting matrix.

In total we used $11$ independent parameters per nucleus, specified on Table~\ref{table:par}. The values of $\chi_\pi$, $\xi_1$=$\xi_2$=$\xi_3$, $\alpha_0$=$\alpha_2$ are kept constant for all eight nuclei and  $\chi_\nu$ is taken the same for the normal and intruder configurations. The variation of $\Delta$ as function of the neutron number is linear, with the same slope as the one suggested in Ref.~\cite{duval-goutte}. Our $\Delta$ values are larger than the ones given in~\cite{duval-goutte} because we are assumming that the intruder configuration originates from the excitation of one proton pair across the $Z=28$ shell gap instead of a proton pair within the same valence space. According to~\cite{sambataro} this linear behavior arises from the monopole contribution to the neutron-proton interaction.

\begin{table}\label{tab}
\begin{center}
\begin{tabular}{cccccccccc}\hline
$A$ \ & $N_\nu$ \ & $\chi_\nu$ \ && $\varepsilon$ \, [MeV] \ && $\kappa$ \, [MeV] \ && $\Delta$ \, [MeV] & \ $e_2$ \,[$e$b] \\\hline
$68$ \ & $4$ \ & $1.45$ \ && $1.40$ ($1.40$) \ && -$0.20$ (-$0.25$) \ && $3.73$ & \ $0.052$ \\
$70$ \ & $5$ \ & $1.40$ \ && $1.40$ ($1.30$) \ && -$0.20$ (-$0.23$) \ && $3.35$ & \ $0.047$ \\
$72$ \ & $\bar 5$ \ & $1.30$ \ && $1.40$ ($1.30$) \ && -$0.21$ (-$0.23$) \ && $2.50$ & \ $0.033$ \\
$74$ \ & $\bar 4$ \ & $1.20$ \ && $1.20$ ($1.10$) \ && -$0.21$ (-$0.23$) \ && $0.94$ & \ $0.032$ \\
$76$ \ & $\bar 3$ \ & $1.12$ \ && $1.00$ ($1.05$) \ && -$0.21$ (-$0.25$) \ && $0.03$ & \ $0.032$ \\
$78$ \ & $\bar 2$ \ & $0.92$ \ && $1.00$ ($1.00$) \ && -$0.23$ (-$0.26$) \ && -$0.98$ & \ $0.032$ \\
$80$ \ & $\bar 1$ \ & $0.85$ \ && $1.00$ ($1.03$) \ && -$0.24$ (-$0.27$) \ && -$1.92$ & \ $0.032$ \\
$82$ \ & $0$ \ & \ && $1.10$ ($1.30$) \ &&  \ && -$3.00$ & \ $0.038$ \\\hline
\end{tabular}
\end{center}
\caption{\label{table:par}Parameters used in this calculation. The bar above the number of neutron-bosons indicates that the bosons correspond to pairs of neutron-holes. The values for the intruder configuration are given in parenthesis. For all the isotopes $N_\pi$=$2$($4$), $\chi_\pi$=-$1.2$(-$1.4$), $\xi_1$=$\xi_2$=$\xi_3$=$0.05$($0.1$), $\alpha_0$=$\alpha_2$=$0.115$ MeV. The effective charges for the normal component, $e_2$, are given in the last column, while for the intruder we took $e_4 = 2 \, e_2$.}
\end{table}

\begin{figure*}
\scalebox{.90}{\includegraphics{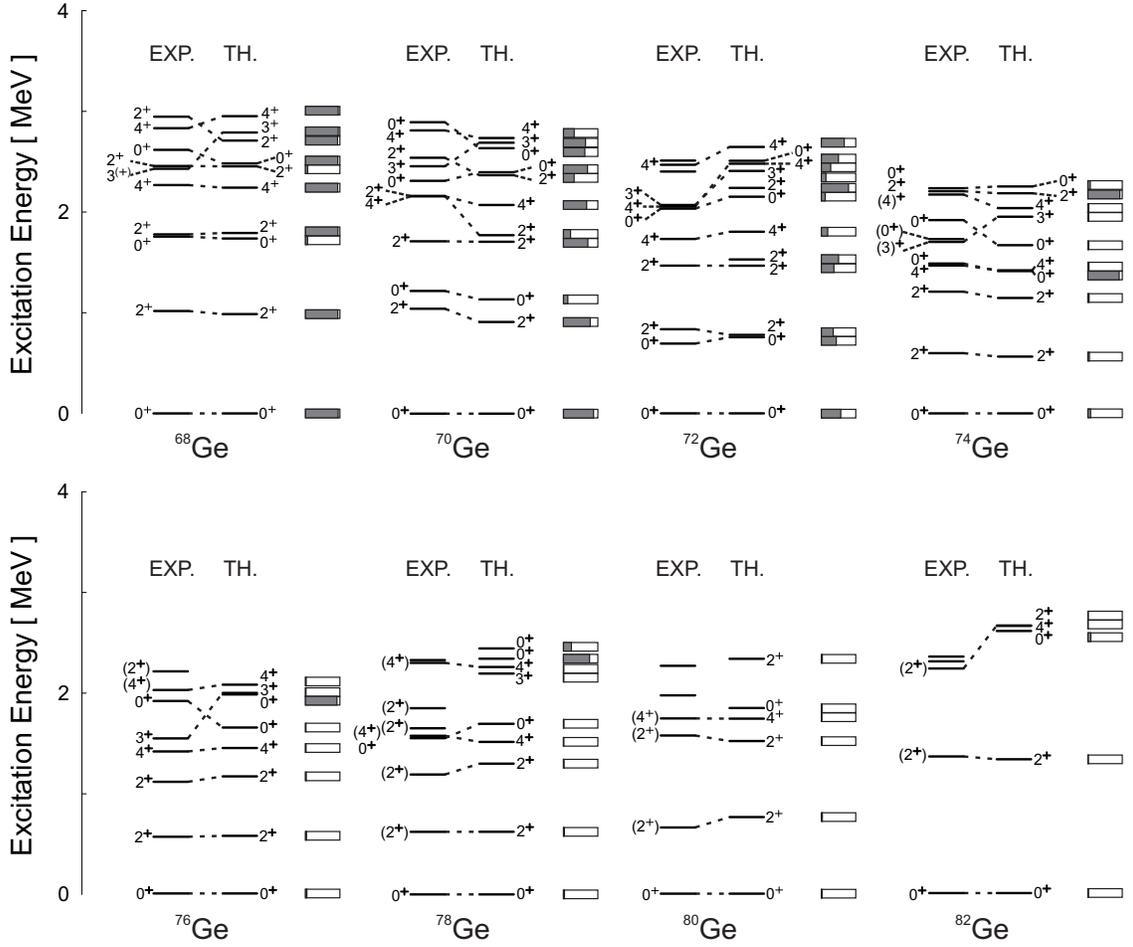}}
\caption{\label{fig:levels} Comparison between the experimental and calculated energy levels for the germanium isotopes. The wave function composition for each theoretical state is given as a normal (gray)-intruder (white) percentage on the right hand side column.}
\end{figure*}

The calculated low-energy levels for the even $^{68-82}$Ge isotopes are shown in Fig.~\ref{fig:levels} together with the experimental data taken from Ref.~\cite{ENSDF}. A satisfactory agreement for the entire isotope chain is obtained. The evolution of the mixing as the neutron number increases, can be seen in Fig.~\ref{fig:levels} by looking at the column next to the theoretical spectra for each isotope. Each horizontal bar gives the eigenfunction composition, the gray portion represents the sum of the square coefficents of the normal components, while the white portion represents the same quantity for the intruder components.

From the Fig.~\ref{fig:levels} one observes a one-to-one correspondence between the experimental and theoretical energy levels for $^{68}$Ge and $^{70}$Ge up to an excitation energy of $\sim 3$ MeV, with the $3_1^+$ state of $^{68}$Ge and the $2_3^+$ state of $^{70}$Ge showing the largest discrepancies. The mixing in the wave functions of $^{68}$Ge is very small and the two configurations appear well separated with the normal (intruder) component been predominant for the low(high) energy levels; for $^{70}$Ge the mixing starts to become important, especially for high energies, while the normal configuration still dominates at energies less than $1$ MeV. For $^{72}$Ge the theoretical calculation yields a $2_3^+$ state which has no experimental counterpart. The existence of such a level has also been suggested by other authors~\cite{duval-goutte} using different theoretical approaches \cite{kumar2}. According to our calculated electromagnetic transitions, $2_3^+$ represents the continuation of the $0_2^+$ band-head. The mixing is maximal for $^{72}$Ge with a nearly $50$\% normal, $50$\% intruder composition of the eigenfunctions. For $^{74}$Ge the two configurations are inverted, and it is now the intruder configuration that dominates the low-energy levels in the spectra, while the normal component becomes important only for higher energy levels. For the isotopes $^{76}$Ge to $^{82}$Ge, the fit of the energy levels is good although there is an increasing lack of experimental information as one moves to the neutron-rich part of the chain. For those isotopes the mixing seems to be less relevant, as there is only one dominant configuration. The extreme case for this situation is the neutron-closed-shell nucleus $^{82}$Ge, that has $N_\nu$=$0$ and therefore a simple IBM-1 calculation is able to reproduce the scarce experimental information available.

In Table \ref{table:emt} we present the most important electric quadrupole transitions between the calculated energy levels for the germanium isotopes. The values are compared with the experimental information available in the literature. The $B(E2)$ values and the quadrupole moments were obtained following the definitions
	\begin{eqnarray}
	B(E2; L \rightarrow L') &=& \frac{1}{2L + 1} \ 
	\vert \langle L' \vert \vert T^{(E2)} \vert \vert L \rangle \vert^2 \, , \\
	&& \nonumber \\
	Q(2_i^+) &=& \left(\frac{32 \pi}{175}\right)^{1/2} \ 
	\vert \langle 2_i^+ \vert \vert T^{(E2)} \vert \vert 2_i^+ \rangle \vert^2 \, ,
	\end{eqnarray}
with the electric quadrupole transition operator given by
	\begin{equation}
	T^{(E2)} = e_2 ( Q_{\pi 2} + Q_{\nu 2} ) + e_4 ( Q_{\pi 4} + Q_{\nu 4} )\, , 
	\end{equation}
being $Q_{\rho j}$, the quadrupole operator defined in equation (\ref{quadrupole}) for the normal ($j=2$) and intruder ($j=4$) configurations. The values of the boson effective charges $e_2$ ($e_4 = 2 e_2$ for all isotopes, following the work of Sambataro and Molnar \cite{sambataro} on the Mo isotopes) were determined by the experimental $B(E2; 2_1^+ \rightarrow 0_1^+)$ values. 

\begin{table}\label{tab2}
\begin{center}
\begin{tabular}{crrrrrrrr}\hline
& \multicolumn{2}{c}{$^{68}$Ge} & \multicolumn{2}{c}{$^{70}$Ge} & \multicolumn{2}{c}{$^{72}$Ge} &  \multicolumn{2}{c}{$^{74}$Ge} \\
& EXP. & TH. \quad \quad & EXP. & TH. \quad \quad & EXP. & TH. \quad \quad & EXP. & TH. \\\hline
$B(E2;2_1^+ \rightarrow 0_1^+)$ \quad \quad & \ $29$(3) & \ $27.2$ \quad \quad & \ $36$ (4) & \ $35.9$ \quad \quad & \ $40$ (3) & \ $39.0$ \quad \quad & \ $60$ (3) & \ $62.2$ \\
$B(E2;2_1^+ \rightarrow 0_2^+)$ \quad \quad &  & \ $4.8$ \quad \quad & \ $13$ (3) & \ $16.5$ \quad \quad & \ $41$ (4) & \ $18.4$ \quad \quad & \ $<7.8$ & \ $3.0$ \\
$B(E2;2_2^+ \rightarrow 2_1^+)$ \quad \quad & \ $0.8$(3) & \ $4.2$ \quad \quad & \ $49.7$ (189) & \ $68.2$ \quad \quad & \ $114$ (12) & \ $59.4$ \quad \quad & \ $99.7$ (203) & \ $91.5$ \\
$B(E2;4_1^+ \rightarrow 2_1^+)$ \quad \quad & \ $22.9$(30) & \ $41.0$ \quad \quad & \ $18.9$ (34) & \ $68.1$ \quad \quad & \ $64.1$ (71) & \ $80.0$ \quad \quad & \ $66.4$ (55) & \ $91.8$ \\\hline
$Q(2_1^+)$ \quad \quad & & \ $4.6$ \quad \quad & \ $3$ (6) & \ $2.1$ \quad \quad & \ -$12$ (8) & \ -$6.1$ \quad \quad & \ -$19$ (2) & \ -$15$ \\
$Q(2_2^+)$ \quad \quad & & \ -$0.3$ \quad \quad &  & $9.8$ \quad \quad & \ $23$ (8) & \ -$19.3$ \quad \quad & \ $26$ (6) & \ $13.0$ \\ \hline \hline
& \multicolumn{2}{c}{$^{76}$Ge} &  \multicolumn{2}{c}{$^{78}$Ge} &  \multicolumn{2}{c}{$^{80}$Ge} & \multicolumn{2}{c}{$^{82}$Ge} \\ 
& EXP. & TH. \quad \quad & EXP. & TH. \quad \quad & EXP. & TH. \quad \quad & EXP. & TH. \\ \hline
$B(E2;2_1^+ \rightarrow 0_1^+)$ \quad \quad& $46$ (3)\ & $52.2$ \quad \quad& $44$ (3)\ & $40.3$ \quad \quad& $28$ (5) \ & $27.6$ \quad \quad& $25$ (5) \ & $27.6$ \\
$B(E2;2_1^+ \rightarrow 0_2^+)$ \quad \quad & $<2.8$ \ & $1.3$ \quad \quad & $.7$  $\left(^{+5}_{-2}\right)$ \ & $3.0$ \quad \quad & \ \quad \quad \quad & $3.5$ \quad \quad & \ \quad \quad \quad \quad & $3.5$ \\
$B(E2;2_2^+ \rightarrow 2_1^+)$ \quad \quad& $74.6$ (96)\ & $73.9$ \quad \quad& $39.6$ $\left(^{+337}_{-139}\right)$\ & $53.2$ \quad \quad& \ & $39.2$ \quad \quad& \ & $39.2$ \\
$B(E2;4_1^+ \rightarrow 2_1^+)$ \quad \quad& $73$ (13)\ & $74.5$ \quad \quad& $>21.8$\ & $57.4$ \quad \quad& \ & $39.0$ \quad \quad& \ & $39.0$ \\\hline
$Q(2_1^+)$ \quad \quad& -$14$ (4)\ & -$15.3$ \quad \quad&\ & -$18.3$ \quad \quad&\ & -$13.6$ \quad \quad&\ & -$0.3$ \\
$Q(2_2^+)$ \quad \quad& $28$ (6)\ & $11.7$ \quad \quad&\ & $11.9$ \quad \quad&\ & $5.2$ \quad \quad&\ & $0.2$ \\ \hline
\end{tabular}
\end{center}
\caption{\label{table:emt}A comparison between the experimental and theoretical $B(E2)$ values and quadrupole moments are given for the Ge isotopes from A = 68 to A = 82. The units of the $B(E2)$ values are given by $10^{-3}\, e^2$ b$^2$ while for the quadrupole moments one uses $10^{-2} \, e$b.}
\end{table}

\section{\label{sec:3}Geometric Interpretation}

To obtain a geometric interpretation of the model we use the coherent states associated to the IBM-2. The most general form of these states is given by~\cite{leviatan}  
	\begin{equation}
      \vert N_{\pi},  N_{\nu},\beta_\pi, \gamma_\pi, \beta_\nu, \gamma_\nu, \phi, \theta, \psi 
	\rangle   = 
	\frac{1}{ \sqrt{(N_\pi)! (N_\nu)!}} \ R (\phi, \theta, \psi) \  
	(\Gamma^{\dagger}_{\pi})^{N_\pi} \ ( \Gamma^{\dagger}_{\nu } )^{N_\nu} 
	\ \vert 0 \rangle \, ,
	\label{coherent}
	\end{equation}
with
      \begin{equation}
	\Gamma^{\dagger}_{\rho} = \left[ s^\dagger_\rho +
	\beta_\rho \cos{\gamma_\rho} d^\dagger_{\rho, 0} 
	+ \frac{1}{\sqrt{2}} \beta_\rho \sin{\gamma_\rho} ( d^\dagger_{\rho, 2} + 
	d^\dagger_{\rho, -2}) \right] / \sqrt{1 + \beta^2_\rho} \, ,  
       \end{equation}
where $ \vert 0 \rangle$ is the boson vacuum, and the Euler angles, $\Omega = (\phi, \theta, \psi)$, define the orientation of the deformation variables $(\beta_\pi, \gamma_\pi)$ for proton-bosons with respect to the corresponding to neutron-bosons $(\beta_\nu, \gamma_\nu)$ . It has been shown~\cite{leviatan} that in the absence of hexadecupole interaction, one can take the Euler angles equal to zero. Using the states~(\ref{coherent}) with $\Omega = 0$, one can evaluate the matrix elements of the normal(intruder) Hamiltonian, $H_N$($H_{N + 2}$). The result for the normal configuration is
	\begin{eqnarray}
	&& E_{N_\pi, N_\nu}( \beta_\pi, \gamma_\pi, \beta_\nu, \gamma_\nu)  
	= \varepsilon \left( \frac{N_\pi \beta^2_\pi}{1 + \beta^2_\pi} + \frac{N_\nu \beta^2_\nu}
	{1 + \beta^2_\nu} \right) 
	+ \frac{2 \kappa N_\pi N_\nu \beta_\pi \beta_\nu}{(1 + \beta^2_\pi)
	(1 + \beta^2_\nu)} 
	\biggl(2 \cos(\gamma_\pi - \gamma_\nu) \nonumber \\
	&-& \sqrt{\frac{2}{7}} \chi_\pi \beta_\pi \cos ( \gamma_\nu + 2 \gamma_\pi) 
	- \sqrt{\frac{2}{7}} \chi_\nu \beta_\nu 
	\cos ( \gamma_\pi + 2 \gamma_\nu) + \frac{1}{7} \chi_\pi \chi_\nu \beta_\nu 
	\beta_\pi \cos ( 2 \gamma_\pi - 2 \gamma_\nu) \biggr) 
	 \nonumber \\
	&+& \xi_2 \frac{N_\pi N_\nu}{(1 + \beta^2_\pi)(1 + \beta^2_\nu)} 
	\left( \left( \beta_\pi - \beta_\nu \right)^2 
	+ 2 \beta_\pi \beta_\nu \left( 1 - 
	\cos ( \gamma_\pi - \gamma_\nu ) \right) \right) \, , 
	\label{surface1}
	\end{eqnarray}
whereas for the intruder, the matrix element denoted as: $\bar E_{N_\pi, N_\nu}( \beta_\pi, \gamma_\pi, \beta_\nu, \gamma_\nu)$, can be obtained from (\ref{surface1}) by replacing the appropriate Hamiltonian parameters and changing $N_\pi$ for $N_\pi + 2$. The geometric interpretation of the IBM-2 with configuration mixing is determined through the diagonalization of the matrix energy surface
	\begin{equation}
	E = \left[ \begin{array}{cc}
	E_{N_\pi, N_\nu}( \beta_\pi, \gamma_\pi; \beta_\nu, \gamma_\nu ) & w( N_\pi, \beta_\pi ) \\
	w( N_\pi, \beta_\pi ) & \bar E_{N_\pi + 2, N_\nu}( \beta_\pi, \gamma_\pi; \beta_\nu, 
	\gamma_\nu )  + \Delta\\
	\end{array} \right] \, , 
	\label{mat1}
	\end{equation}
where $w(N_\pi, \beta_\pi)$ denotes the matrix element of the mixing Hamiltonian (\ref{hmix}) in the coherent states (\ref{coherent}), with $\Omega =0$. The explicit form of this term is the following
	\begin{equation}
	w(N_\pi, \beta_\pi) = \frac{\sqrt{( N_\pi + 1 ) ( N_\pi + 2 )}}{ 1 + \beta^2_\pi} \ 
	\left( \alpha_0 + \frac{\alpha_2}{\sqrt{5}} \ \beta_\pi^2 \right)\, .
	\end{equation}
The solution of the eigenvalue problem of (\ref{mat1}) leads to two energy surfaces 
	\begin{eqnarray}
	E_{\pm}( \beta_\pi, \gamma_\pi; \beta_\nu, \gamma_\nu, \Delta ) &=& E_{N_\pi, N_\nu}
	( \beta_\pi, \gamma_\pi; \beta_\nu, \gamma_\nu ) + g( \beta_\pi, \gamma_\pi; 
	\beta_\nu, \gamma_\nu, \Delta ) \nonumber \\
	& \pm & \sqrt{g^2( \beta_\pi, \gamma_\pi; 
	\beta_\nu, \gamma_\nu, \Delta ) + w^2(N_\pi, \beta_\pi)}
	\label{sol12}
	\end{eqnarray}
where 
	\begin{equation}
	g( \beta_\pi, \gamma_\pi; \beta_\nu, \gamma_\nu, \Delta ) = \frac{1}{2}
	\biggl( \bar E_{N_\pi + 2, N_\nu}( \beta_\pi, \gamma_\pi; \beta_\nu, \gamma_\nu ) - 
	E_{N_\pi, N_\nu}( \beta_\pi, \gamma_\pi; \beta_\nu, \gamma_\nu ) + \Delta \biggr) \, . 
	\end{equation}
The corresponding eigenfunctions are 
	\begin{equation}
	X_{+} = \frac{1}{\sqrt{2R}} \ \left[
	\begin{array}{c}
	\sqrt{R-1} \\ \sqrt{R+1}
	\end{array} \right] \, , \qquad
	X_{-} = \frac{1}{\sqrt{2R}} \ \left[
	\begin{array}{c}
	-\sqrt{R+1} \\ \sqrt{R-1}
	\end{array} \right] \, , 
	\end{equation}
with $R = \sqrt{1 + \left(w(N_\pi, \beta_\pi)/g( \beta_\pi, \gamma_\pi; \beta_\nu, \gamma_\nu, \Delta )\right)^2}$. From the equation (\ref{surface1}) one can notice that by taking $\beta_\pi = \beta_\nu \rightarrow \beta$ and $\gamma_\pi = \gamma_\nu \rightarrow \gamma$ the contribution of the Majorana interaction to the energy surface is zero. Under this condition the other terms in~(\ref{surface1}) reduce to the energy surface associated to the IBM-1
	\begin{equation}
	E(N, \beta, \gamma) = \frac{\bar \varepsilon N \beta}{1 + \beta^2} + 
	\frac{N(N-1)}{(1 + \beta^2)^2} \ \left( a_1 \beta^4 + a_2 \beta^3 \cos 3\gamma
	+ a_3 \beta^2 \right) \, ,
	\end{equation}
for the diagonal terms of (\ref{mat1}), with
	\begin{eqnarray}
	\bar \varepsilon &=& \varepsilon + \kappa \frac{2 N_\pi N_\nu}{N} \, , \qquad \qquad
	\qquad \qquad a_1 = \frac{2 \kappa N_\pi N_\nu}{N(N-1)}\
	\left( -1 + \frac{\chi_\pi \chi_\nu}{7}\right) \, , \\
	a_2 &=& - \sqrt{\frac{2}{7}} \ \frac{2 \kappa N_\pi N_\nu}{N(N-1)} (\chi_\pi + \chi_\nu) \, , 
	\qquad 
	a_3 = \frac{2 \kappa N_\pi N_\nu}{N(N-1)}\, ,
	\end{eqnarray}
and 
	\begin{equation}
	w (N, \beta) = \sqrt{\frac{(N_\pi + 1)(N_\pi + 2)}{1 + \beta^2}} \left( \alpha_0 + 
	\frac{\alpha_2}{\sqrt{5}} \beta^2 \right) \, ,
	\end{equation}
for the non-diagonal terms. Thus one concludes that the condition on $\beta_\rho$ and $\gamma_\rho$ mentioned above is equivalent to the projection of the IBM-2 to the IBM-1~\cite{scholten}.

The first step followed in the study of the geometry associated to the IBM-2 plus configuration mixing for the Ge isotopes, was to consider the condition  $\beta_\pi = \beta_\nu \rightarrow \beta$, $\gamma_\pi = \gamma_\nu \rightarrow \gamma$. To convince ourselves that such consideration makes sense, we performed a numerical calculation taking a large strength of the Majorana interaction. The result shows that indeed the wave functions as well as the energy levels associated to the ground band are almost not affected.

\begin{figure*}
\scalebox{.85}{\includegraphics{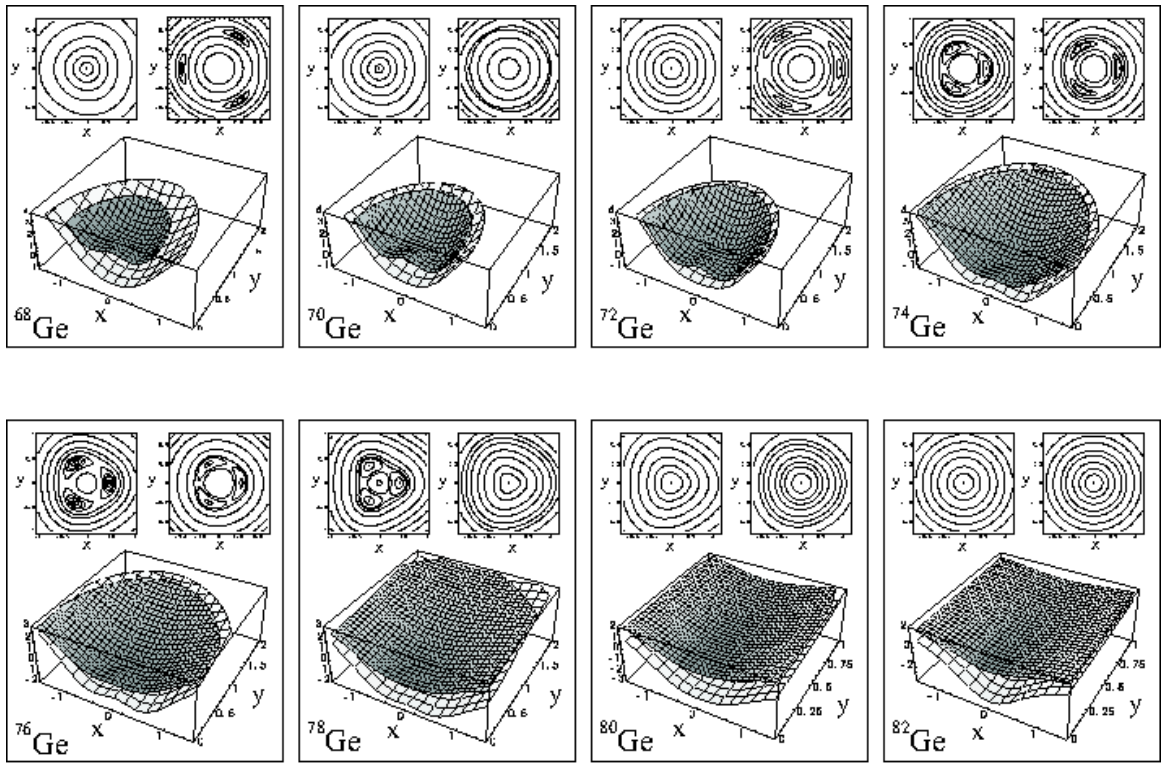}}
\caption{\label{fig:ES} Energy surfaces associated to the ground (white) and excited (gray) bands are shown together with their corresponding contour plots for each one of the Ge isotopes. $x = \beta \cos \gamma$ and $y = \beta \sin \gamma$. The left contour plot for each isotope is associated to the ground band while the right one belongs to the excited band. The dots indicate the deepest contour level of each energy surface.}
\end{figure*}

The energy surfaces obtained for the Ge isotopes are presented in Fig. \ref{fig:ES}. We display both the minimum and excited energy surfaces (see equation (\ref{sol12})) as $3$D-surfaces, together with their corresponding contour plots. One can see that for $^{68}$Ge there is coexistence between a spherical shape for the ground band and an oblate shape for the excited band; in the case of $^{70}$Ge there is a coexistence between spherical and $\gamma$-unstable deformations; for $^{72}$Ge, the most mixed isotope, the lower energy band is spherical while excited energy levels are prolate. According to this interpretation, a shape transition occurs in $^{74}$Ge, where one gets two different prolate shapes for the ground and excited bands; for $^{76}$Ge a similar behavior than the one associated to $^{74}$Ge is found. Finally, there is a gradual evolution towards spherical shapes for the neutron-rich nuclei, in $^{78}$Ge the coexistence is between a prolate ground band and an spherical excited band; in $^{80}$Ge and $^{82}$Ge both energy surfaces are spherical.

\section{Summary}

In summary, we have presented a configuration mixing calculation for the even-even Ge isotopes including the radioactive isotopes $^{78,80,82}$Ge. The good agreement between the theoretical and the experimental energy spectra, $E2$ transitions and quadrupole moments, supports the hypothesis that for light germanium isotopes ($A=68-76$) the interplay of two configurations determines the low-energy structure of the nuclei. In this calculation we have assumed that the intruder configuration arises from the two-proton two-hole excitation across the $Z=28$ shell gap. Our extrapolation to heavier isotopes ($A=78-82$) suggets that the configuration mixing is less important. However a definitive conclusion requires more experimental information about these nuclei. By means of a matrix formulation a geometric interpretation of the IBM-2 with configuration mixing was introduced. According to this each nucleus is described as a superposition of two energy surfaces that carry information about the equilibrium deformation parameters. It is shown that the projection $\beta_\pi = \beta_\nu \rightarrow \beta$ and $\gamma_\pi = \gamma_\nu \rightarrow \gamma$  of these two energy surfaces reduces to the geometric interpretation of the IBM-1 with configuration mixing. For the Ge isotopes, it is found that increasing the strength of the Majorana interaction does not affect significantly the energies and $B(E2)$ values of the ground state bands, justifying the use of IBM-1 projection to analyze the geometry. One finds that the shape of the ground band evolves from spherical in $^{68,70,72}$Ge to prolate in $^{74,76,78}$Ge with a shape phase transition from spherical to prolate nuclei occurring between $^{72}$Ge and $^{74}$Ge. The energy surfaces characterize the ground and excited bands of the Ge isotopes which have in general different shapes and an orthogonal composition of the normal ($N$) and intruder ($N+2$) coherent states.

\begin{acknowledgments}
This work was partially supported by CONACyT. Oak Ridge National Laboratory is managed by UT-Battelle, LLC, for the U.S. DOE under the Contract DE-AC05-00OR22725.
\end{acknowledgments}

\end{document}